\documentclass[10pt,twocolumn,superscriptaddress,prl,aps]{revtex4-1}

\usepackage{amsmath}    
\usepackage{graphicx}   
\usepackage{verbatim}   
\usepackage{color}      
\usepackage{subfigure}  
\usepackage{hyperref}   
\usepackage{units}
\usepackage{braket}
\usepackage[dvipsnames]{xcolor}
\usepackage{ulem}
\begin{document}

\title{Measurement of spectral functions of ultracold atoms in disordered potentials}

\author{Valentin V. Volchkov} 
\affiliation{Laboratoire Charles Fabry, Institut d'Optique, CNRS, 2 Avenue Augustin Fresnel 91127 Palaiseau cedex, France}
\affiliation{Max-Planck-Institute for Intelligent Systems, Spemannstrasse 34, 72076 T\"ubingen, Germany}

\author{Michael Pasek}
\altaffiliation[Present address: ]{Coll\`ege de France, 11 place Marcelin Berthelot, 75005 Paris, France}
\affiliation{Laboratoire Charles Fabry, Institut d'Optique, CNRS, 2 Avenue Augustin Fresnel 91127 Palaiseau cedex, France}
\affiliation{Laboratoire Kastler Brossel, UPMC-Sorbonne Universit\'es, CNRS, ENS-PSL Research University, Coll\`{e}ge de France, 4 Place Jussieu, 75005 Paris, France}

\author{Vincent Denechaud} 
\affiliation{Laboratoire Charles Fabry, Institut d'Optique, CNRS, 2 Avenue Augustin Fresnel 91127 Palaiseau cedex, France}
\affiliation{SAFRAN Sensing Solutions, Safran Tech, Rue des Jeunes Bois, Ch\^{a}teaufort CS 80112, 78772 Magny-les-Hameaux, France}

\author{Musawwadah Mukhtar} 
\affiliation{Laboratoire Charles Fabry, Institut d'Optique, CNRS, 2 Avenue Augustin Fresnel 91127 Palaiseau cedex, France}

\author{Alain Aspect} 
\affiliation{Laboratoire Charles Fabry, Institut d'Optique, CNRS, 2 Avenue Augustin Fresnel 91127 Palaiseau cedex, France}

\author{Dominique Delande}
\affiliation{Laboratoire Kastler Brossel, UPMC-Sorbonne Universit\'es, CNRS, ENS-PSL Research University, Coll\`{e}ge de France, 4 Place Jussieu, 75005 Paris, France}

\author{Vincent Josse} 
\email[Corresponding author: ]{vincent.josse@institutoptique.fr}
\affiliation{Laboratoire Charles Fabry, Institut d'Optique, CNRS, 2 Avenue Augustin Fresnel 91127 Palaiseau cedex, France}

\pacs{}

\date{\today}

\begin{abstract}
We report on the measurement of the spectral functions of non-interacting ultra-cold atoms
in a three-dimensional disordered potential resulting from an optical speckle field. 
Varying the disorder strength by two orders of magnitude, we  observe the crossover from the ``quantum'' perturbative regime of low disorder to the ``classical'' regime at higher disorder strength, and find an excellent agreement with numerical simulations. 
The method relies on the use of state-dependent disorder and the controlled transfer of atoms to create well-defined energy states.
This opens new avenues for experimental investigations of three-dimensional Anderson localization.

\end{abstract}

\maketitle

\textit{Introduction.---}
The spectral function provides essential information on the energy-momentum relation of one-particle excitations in complex systems.
This relation takes a non-trivial form in the presence of random scatterers or inter-particle interactions \cite{bruus}.
The direct measurement of the spectral function via angle-resolved photoemission spectroscopy (ARPES) \cite{Damascelli04} in strongly correlated electronic systems has led to significant progress in the understanding of high-$T_\mathrm{c}$ superconductivity \cite{Damascelli03RMP}.
More recently, the ability to measure and exploit spectral functions in ultracold atomic systems has also been widely demonstrated, for instance using radio-frequency spectroscopy \cite{Dao07,Stewart08} to reveal the presence of a pseudogap in strongly interacting Fermi gases \cite{Gaebler10,feld2011}, or to probe the Mott insulator and superfluid regimes of interacting Bose gases in periodic lattices using Bragg spectroscopy \cite{clementfabbri,ernst2010,fabbri2012}.

\begin{figure}[!b]%
	\centering
		\includegraphics[width=1\columnwidth]{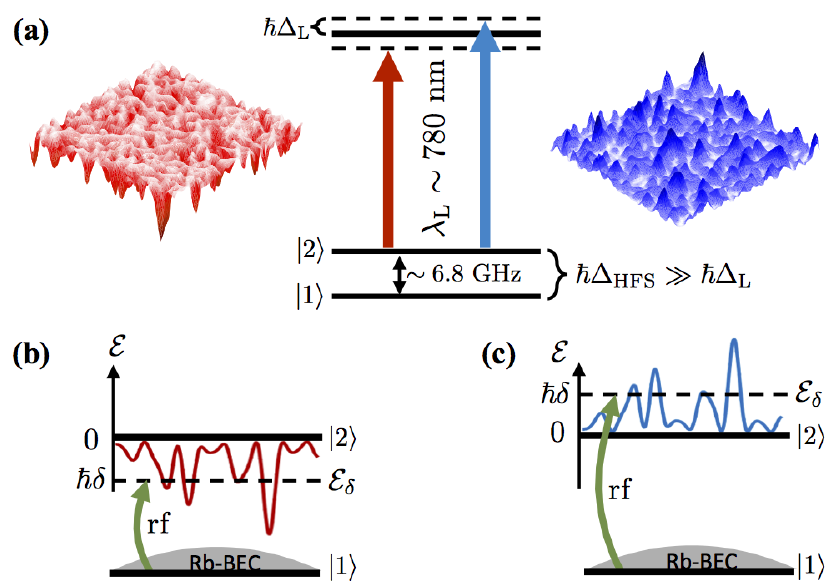}%
		\caption{\label{fig:scheme} Measurement scheme of the spectral function using a state-dependent disordered potential.
	(a) A near-resonant laser speckle field of detuning $\Delta_\mathrm{L}$ creates either an attractive (red-detuned, $\Delta_\mathrm{L}<0$) or repulsive (blue-detuned, $\Delta_\mathrm{L}>0$) disordered potential on atoms in internal state $\ket{2}$, while the disordered potential experienced by atoms in internal state $\ket{1}$ is negligible, since $\Delta_\mathrm{HFS}\gg \Delta_\mathrm{L}$.
	(b) and (c) A radio-frequency field at frequency $\Delta_\mathrm{HFS}+\delta$ transfers a small fraction of atoms in a BEC in state $\ket{1}$ to the state $\ket{2}$. The transfer rate measured in this experiment is proportional to the spectral function, according to the Fermi golden rule [see Eq.~\eqref{Ak_def}].}
\end{figure}

In disordered systems, the knowledge of the spectral function is also crucial, from the search of gapless excitations in the Bose glass phase in presence of interactions (see e.g.~\cite{Derrico2014}) to the precise investigation of the Anderson quantum phase transition for non-interacting particles~\cite{anderson1958,abrahams2010}. 
In the Anderson localization problem, the spectral function is not only a basic ingredient used in theoretical approaches to predict the position of the mobility edge (the critical energy of the transition)~\cite{Vollhardt1992self} but it is also used as a resource to extract an approximate value of the mobility edge from experimental observations \cite{kondov2011,jendrzejewski2012a,semeghini2014}.
Significant discrepancies observed between the experiments and theoretical analyses~\cite{piraud2012,fratini2015,pasek2016anderson} render the precise measurement of these spectral functions yet more desirable.

In this letter, we report on the direct measurement of the spectral function at quasi-null momentum of non-interacting ultracold atoms in continuous three-dimensional (3D) laser speckle disordered potentials. We explore a large range of disorder strengths, from the so-called ``quantum" regime of weak disorder (see e.g.~\cite{kuhn2007}), where the spectral function is a narrow function whose width gives the inverse lifetime of the initial momentum state, to the so-called ``classical" regime of strong disorder, where atoms can be described by a semi-classical wavefunction and spectral functions converge towards the probability distribution of the disorder~\cite{trappe2015,PratPRA16}. The measurements are done both with an attractive (red-detuned) and a repulsive (blue-detuned) laser speckle disorder, the latter case being particularly important since most experimental studies of Anderson localization of ultra-cold atoms have been done in that configuration. Numerical calculations are in excellent agreement with the experimental results, not only in the marginal regimes of weak and strong disorder, but also in the crossover in-between where finding accurate expressions is a theoretical challenge~\cite{yedjour2010,piraud2013,pasek2015cpa}.

The method is based on a radio-frequency (rf) transfer of atoms at rest in an atomic internal state $\ket{1}$ insensitive to the disorder, to a final internal state $\ket{2}$ sensitive to the disordered potential (see Fig.~\ref{fig:scheme})~\cite{clement2006}. The transfer allows us to selectively populate eigenstates of the random potential around the resonant energy $E_\mathrm{f}=E_\mathrm{i}+\hbar\omega$ set by the rf frequency $\omega$ (here $E_\mathrm{i,f}$ corresponds to the \textit{total} energy of the initial and final states). Due to the finite energy resolution of the transfer, energy levels in the disorder behave as an effective continuum, whose density of states $\rho$ is equal to the density of states averaged over disorder realizations~\cite{supp}.  According to the Fermi golden rule, one can thus define a transfer rate $\Gamma$, proportional to the squared modulus of the transition amplitude from the initial state $\ket{1}$ to the targeted final states, which is directly linked to the spectral function of the disordered potential [see Eq.~\eqref{Ak_def} below].
We start indeed with atoms in a dilute Bose-Einstein condensate (BEC) in a shallow trap, whose wavefunction is very close to a null momentum state $\ket{\mathbf{k}\!=\!0}$ such that the total energy of the initial state can be taken equal to the internal energy $E_1$~\cite{supp}. The external energy of the final states is then given by $\mathcal{E}_\delta=\hbar\delta$, where $\delta=\omega-\Delta_\mathrm{HFS}$ is the rf detuning from the bare resonant frequency corresponding to the hyperfine splitting between the respective internal energies $\Delta_\mathrm{HFS}/2\pi = (E_2-E_1)/h\simeq 6.8$~GHz (see Fig.~\ref{fig:scheme}). The rf transfer being associated with a negligible momentum change, the transfer rate from state $\ket{1}$ to $\ket{2}$ is thus proportional to the spectral function $A(\mathcal{E}_\delta,\mathbf{k}=0)$:
 \begin{eqnarray}
	\label{Ak_def}
	\Gamma \propto  A(\mathcal{E}_\delta,\mathbf{k}=0) &=&\overline{\sum_\alpha |\langle \mathbf{k}\!=\!0 | \psi_\alpha \rangle |^2\ \delta(\mathcal{E}_\delta-\mathcal{E}_\alpha)} \nonumber  \\
	&\sim& \overline{ |\langle \mathbf{k}\!=\!0|\psi_{\delta}\rangle|^2} \ \rho(\mathcal{E}_\delta)  
\end{eqnarray}
Here $\ket{\psi_\alpha}$ corresponds to the eigenstate of energy $\mathcal{E}_\alpha$ and $\overline{\dotsb}$ denotes the averaging over disorder realizations. One can thus determine the spectral function by measuring the transfer rate as a function of the rf detuning $\delta$.

\begin{figure*}[ht!]%
	\centering
		\includegraphics[width=0.9\textwidth]{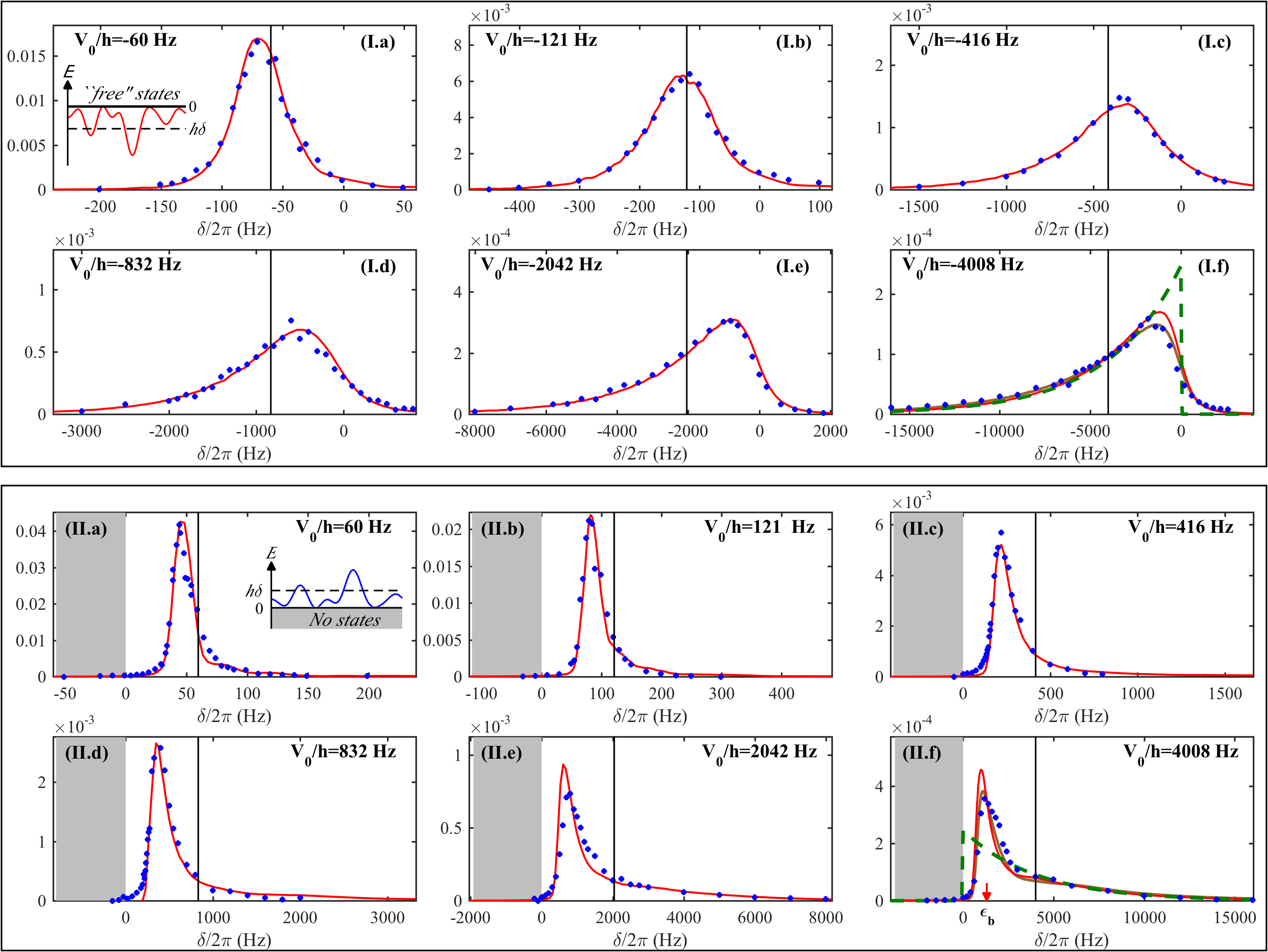}%
	\caption{\label{fig:resultats} Measured (blue dots) and numerically calculated (red solid lines) spectral functions $A(\mathcal{E}_\delta=\hbar\delta,\mathbf{k}\!=\!0)$ of atoms in  attractive (Panel I) or repulsive (Panel II) disordered potentials with various amplitudes. Raw numerical results have been convolved by the experimental resolution function, yielding only minor corrections. The solid brown lines in panels (I.f)  and (II.f) are the results of  numerical computations taking into account the residual effect of disorder in the initial state $\ket{1}$~\cite{supp}. In each panel, the black vertical lines indicate the average value $V_0/h$ of the disorder. The small arrow in panel (II.f) indicates the estimated position of the average ground state energy in local minima, $\mathcal{E}_\mathrm{b}/h=1.3~\mathrm{kHz}$ (see text). Insets in panels (I.a) and (II.a) illustrate the disorder potential for the corresponding configuration. The probability distribution $P(V)$ of the speckle potential is represented as a dashed green curve in panel (I.f) and (II.f) for comparison.} 
\end{figure*}

\textit{Experiment.---}
An original feature of the experiment is the realization of a state-dependent disordered potential significant only for the state $\ket{2}$. As sketched in Fig.~\ref{fig:scheme}(a), we use a laser close to the hyperfine transition $F=2 \leftrightarrow F'=3$ around the $D_2$ line of rubidium at wavelength $\lambda_L\sim780.24~\textrm{nm}$. 
Tuning the laser at $\Delta_\mathrm{L}/2\pi \simeq \pm \, 80\,\mathrm{MHz}$ from the resonance, we create respectively an attractive (red-detuned) or repulsive (blue-detuned) potential for the $F=2$ state, while the effect is 100 times smaller on the $F=1$ state since $\Delta_\mathrm{HFS}\gg \Delta_\mathrm{L}$~\cite{supp}. The laser speckle is obtained by passing the laser beam through a diffusive plate~\cite{goodman2007}, which yields a well-characterized disordered potential $V(\mathbf{r})$~\cite{clement2006,supp}. The attractive and repulsive cases differ by their probability distribution $P(V)=|V_0|^{-1}\exp{\left[-V/V_0\right]}\; \Theta(V/V_0)$, with $\Theta$ the unit step function, the average value $V_0$ of the potential being respectively negative or positive. The amplitude of the disorder $|V_0|$ is proportional to the laser intensity, and can be varied over two orders of magnitude (see Figure~\ref{fig:resultats}).

In order to obtain a sharp resonance for the $\ket{1}\leftrightarrow  \ket{2}$ transition, we use the two ``clock states'' $ \ket{F=1,m_F=-1}\equiv \ket{1}$ and $\ket{F=2,m_F=+1}\equiv\ket{2}$, whose energy difference is insensitive (at first order) to magnetic fluctuations at the ``magic'' magnetic  field of $B_0=3.23~\textrm{G}$, which we impose on the atoms. 
The result is a resonance of width about 10~Hz. Note that since the two states have an angular momentum difference $\Delta m_F =2$, we use a two-photon rf transition, involving a microwave and a rf field~\cite{supp}.

The experiment starts with the realization of a $^{87}$Rb-BEC of about $n_1=2\times10^5$ atoms in the state $\ket{1}$. At the same time, the disordered potential for state $\ket{2}$ is turned on. The microwave and  rf fields driving the $\ket{1}  \leftrightarrow \ket{2}$ transition are then applied for a time duration $t$. The rf coupling is weak enough such that the transfer rate $\Gamma$ can be calculated via the Fermi golden rule as written in Eq.~\eqref{Ak_def}~\cite{Moy:1999,Jack:1999,Gerbier:2001,grynberg2010introduction}. The duration $t$ is chosen short enough, i.e. $\Gamma t\ll 1$, such that only a small fraction of atoms is transferred (a few percents at most). At this short time scale, the population in state $\ket{2}$ grows linearly with time as $n_2(t)\simeq n_1(0)\, \Gamma \, t$ and the transfer rate is directly obtained by counting the atoms via fluorescence imaging. The spectral function $A\left(\mathcal{E}_\delta=\hbar \delta,0\right)$ is finally obtained by repeating the measurement at various values of the detuning $\delta$. In practice, we adapt the energy resolution,  $  \Delta \mathcal{E}=\hbar / t$, to the typical energy span of the spectral function for each disorder amplitude, so that it does not affect the observed profile.

\textit{Numerical calculations.---} 
The experimental results are compared to the results of numerical calculations that take into account the detailed statistical properties of the laser speckle used in the experiments (see~\cite{supp}). The calculations are based on the temporal representation of the spectral function 
\begin{equation}
\label{Ak_temp}
A(\mathcal{E}_\delta,\mathbf k) = \frac{1}{\pi\hbar}\,\mathrm{Re}\,\int_0^{\infty} \overline{\langle \mathbf k|\mathrm{e}^{-\mathrm{i} Ht/\hbar}|\mathbf k \rangle}  \mathrm{e}^{\mathrm{i} \mathcal{E}_\delta t/\hbar}\ \mathrm{d} t,
\end{equation}
which amounts to evaluating the (disorder-averaged) scalar product between the initial plane-wave excitation $\ket{\mathbf k}$
and the time-evolved state $\exp(-\mathrm{i} Ht/\hbar)|\mathbf k \rangle$, with $H$ the disordered Hamiltonian. Our time-propagation algorithm uses an iterative scheme based on the expansion of the time-evolution operator in series of Chebyshev polynomials of the Hamiltonian~\cite{roche1997,fehske2009}.

\textit{Results.---} Figure~\ref{fig:resultats} shows the measured spectral functions $A\left(\mathcal{E}_\delta,0\right)$ as well as the results of their numerical calculations, for the cases of  attractive  (Panel I), and repulsive (Panel II) disordered potentials with amplitudes $|V_0|$ ranging from 60~Hz to 4~kHz. The area under the experimental curves is normalized in order to allow for a direct comparison with numerical calculations~\cite{supp}. The disorder strength has been precisely calibrated by adjusting the experimental and numerical curves of panel (I.b), leading to a 14$\%$ correction of  the  amplitude estimated from photometric measurements. This correction factor is then applied to all other measurements. The agreement is excellent over the whole range of disorder amplitudes. 
Note that, in contrast with numerical calculations, no disorder-averaging was necessary for the experimental data.
This is due both to the finite experimental energy resolution that provides an effective averaging over many energy states, and to the very large expansion of the initial BEC that ``samples'' efficiently the disordered potential.

In the attractive case (Fig.~\ref{fig:resultats}, Panel I), we observe a smooth crossover from the weak disorder regime [Panel (I.a)], where the spectral function is relatively narrow, symmetrical and centered closed to the averaged disorder amplitude $V_0$, to the strong disorder regime [Panel (I.f)] where it becomes strongly asymmetrical. These two marginal regimes can be understood by introducing an important energy scale of the problem, the correlation energy, $\mathcal{E}_\sigma = \hbar^2/m(\sigma_\perp^2 \sigma_\parallel)^{2/3}$~\cite{pasek2016anderson} associated with the finite spatial correlations lengths of the disordered potential. Here $m$ is the atomic mass, while $\sigma_\perp$ and $\sigma_\parallel $ are respectively the transverse and longitudinal correlation lengths of the anisotropic laser speckle intensity~\cite{supp}. For our experiment  $\sigma_{\perp} \sim 0.306~\mu\textrm{m}$ and  $\sigma_\parallel \sim 1.45~\mu\textrm{m}$ leading to $\mathcal{E}_\sigma/h \approx 441~\mathrm{Hz}$.

\begin{figure}[!t]%
	\centering
		\includegraphics[width=0.9\columnwidth]{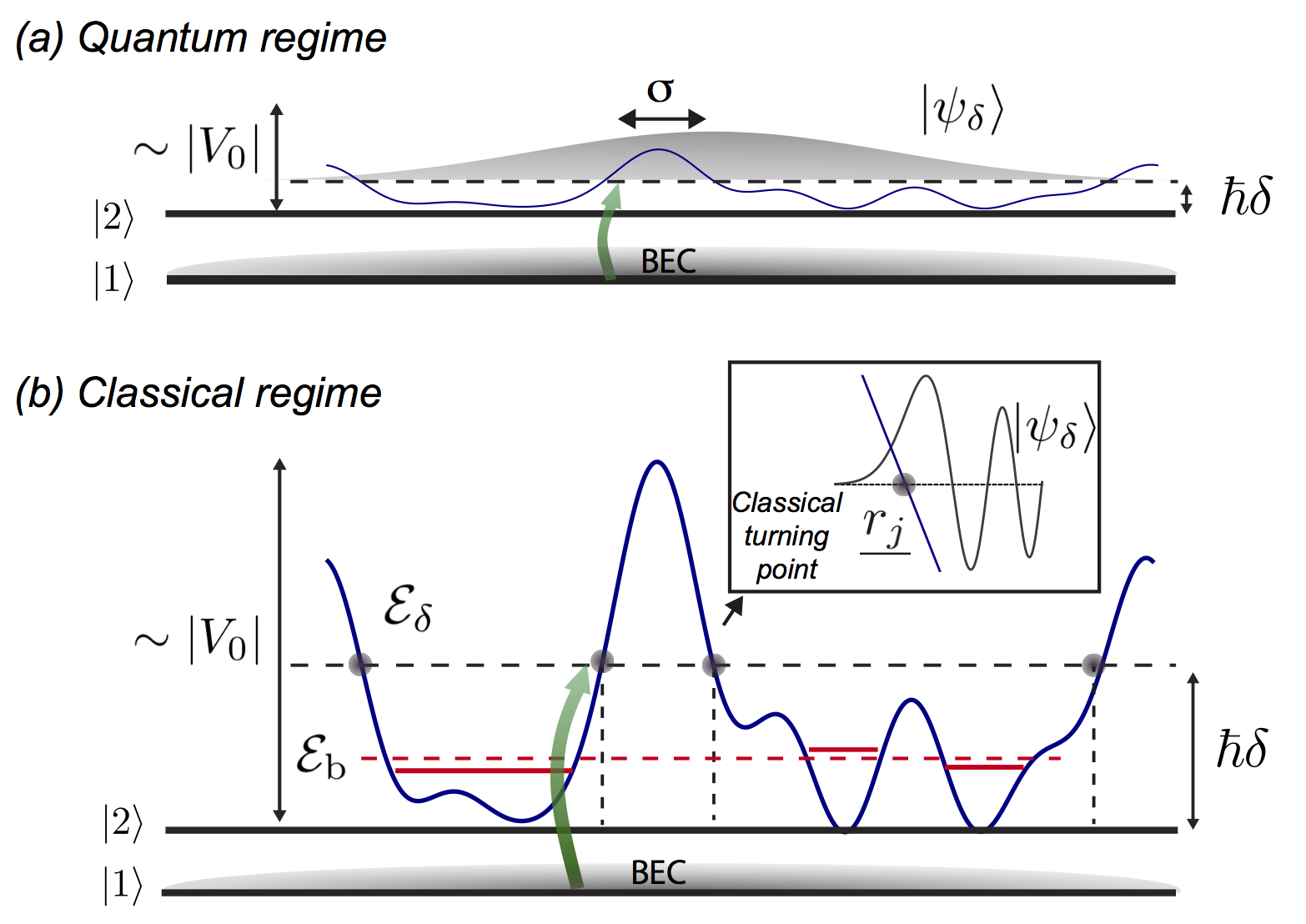}%
	\caption{\label{fig:phys} Schematics of the two marginal regimes of spectral functions of non-interacting atoms in disordered potentials (here shown for the blue-detuned case). (a): Weak disorder $|V_0|\ll \mathcal{E}_\sigma$:  ``quantum'' regime , (b): Strong disorder $|V_0|\gg \mathcal{E}_\sigma$:  ``classical'' regime. $\mathcal{E}_\mathrm{b}$ corresponds to the average energy of the ground states in local potential minima (see text) for blue-detuned laser speckle disordered potential.}
\end{figure}

In the quantum regime [$|V_0|\ll \mathcal{E}_\sigma$, see Panel (I.a)], the amplitude of the disordered potential is too small to support bound states on the typical size $\sigma=(\sigma_\perp^2 \sigma_\parallel)^{1/3}$ of a speckle grain. Atoms with an energy of the order of $|V_0|$ have a large de Broglie wavelength compared to $\sigma$ and their wavefunction extends over many speckle grains [see Fig.~\ref{fig:phys}(a)]. This leads to a smoothing of the disordered potential (see e.g.~\cite{Shklovskii2008,shapiro2012}), whose rescaled effective amplitude corresponds to the width of the spectral function. Alternatively, a perturbative approach of scattering allows us to interpret this width as the inverse lifetime $\hbar / \tau_\mathrm{S}$, where $\tau_\mathrm{S}$ is the elastic scattering time, of the initial state $\ket{\mathbf{k}=0}$~\cite{akkermans2007}. This approach predicts a Lorentzian shape for the spectral function, with a width $\sim \pi V_0^2/\mathcal{E}_\sigma$~\cite{kuhn2007,piraud2013,shapiro2012}. This explains the quasi-Lorentzian shape shown in panel (I.a).

In the classical regime [$|V_0|\gg \mathcal{E}_\sigma$, see panel (I.f)] the situation is the opposite: atoms with an energy of the order of $|V_0|$ have a de Broglie wavelength small compared to $\sigma$. The corresponding wavefunctions have short spatial oscillations, except around the turning points $r_j$ selected by the resonance condition $V(r_j)=\hbar \delta$, where atoms bounce classically on the disordered potential [see Fig.~\ref{fig:phys}(b)]. The overlap with the uniform initial state $\ket{\mathbf{k}\!=\!0}$ is thus negligible except at these positions (the so-called  ``Franck-Condon principle''). The transfer rate - or equivalently the spectral function - is then a probe of the points where $V=\hbar \delta$, i.e. the probability distribution $P(V)$. This property was used in Ref.~\cite{clement2006} to estimate the disorder amplitude $V_0$. Alternatively it can be retrieved using the formal expression of the spectral function $A(\mathcal{E},\mathbf{k})=\overline{\bra{ \mathbf{k}}  \delta (\mathcal{E}-H) \ket{ \mathbf{k}}} $~\cite{supp}. Neglecting the kinetic energy term when $|V_0|\gg \mathcal{E}_\sigma$, it yields directly $A(\mathcal{E},\mathbf{k}\!=\!0)=P(V)$ (see e.g.~\cite{trappe2015}). Consistently, we observe that the spectral function converges at strong disorder towards the probability distribution of the speckle potential [dashed green curve in panel (I.f)]. However, the de Broglie wavelength of atoms remains large around $|\mathcal{E}_\delta|\sim0$, so that the spectral function smoothes-out the sharp discontinuity of the potential distribution.

If we consider now the repulsive case (Panel II), the potential distribution is bounded from below with no state in the negative energy range (gray area). This has two consequences. First the spectral function is strictly zero for negative energy. Second, in the strong disorder regime, the
low energy states that are supported by local minima of the disordered potential lead to an accumulation of states around the averaged ground state harmonic oscillator energy 
$\mathcal{E}_\mathrm{b}= \sqrt{V_0 \mathcal{E}_\sigma}$~\cite{trappe2015,PratPRA16}. This results in a pronounced and narrow peak in the spectral function, which is clearly visible in panel (II.f) around the energy $\mathcal{E}_\mathrm{b}$ (within 30$\%$). While the qualitative explanation was given in the strong-disorder limit, let us note that the peak is present in all the spectral functions shown in panel II. At the lowest disorder amplitude (II.a), it results in a very narrow spectral function, significantly narrower than for the attractive case (I.a) and far from perturbative predictions. This behavior is fully consistent with the strong departure from the perturbative Born prediction observed in the direct measurement of the elastic scattering time $\tau_S$ (which is related to the inverse of the spectral function's width as discussed above) for the same disorder configuration~\footnote{In preparation}. These observations emphasize the difficulties encountered when approximate theories of Anderson localization use perturbative expressions of the spectral function as a resource (see e.g. Refs.~\cite{kuhn2007,yedjour2010,piraud2013}).

\textit{Conclusion.---} We have demonstrated a method that uses a state dependent disordered potential to probe the spectral functions of ultracold atoms in 3D laser speckle potentials. This allowed us to study the crossover from the quantum to the classical regime, the behavior being significantly different for red-detuned or blue-detuned laser speckles. In the latter case, a pronounced peak attributed to lowest bound states in potential minima is observed, resulting in strong deviations from what we would expect using a weak-scattering perturbative approach. The present method, that yields the spectral function around zero momentum, could easily be generalized to finite values of $\mathbf{k}$ by, for instance, using stimulated Raman transitions effected by two laser beams whose angle allows one to select the desired value of $\mathbf{k}$~\cite{kozuma1999,supp}. Besides the measurement of the spectral functions, a key feature of the presented method is the controlled transfer of atoms to well-defined energy states in the disorder, the targeted energy being chosen by the resonance condition.  It opens the possibility to probe the 3D Anderson transition, via a subsequent wavepacket expansion, with an unprecedented energy resolution compared to earlier experimental attempts~\cite{kondov2011,jendrzejewski2012a,semeghini2014}. The interest ranges from the precise location of the mobility edge in such spatially continuous disordered potentials~\cite{fratini2015,pasek2016anderson} to the investigation of the critical regime~\cite{Slevin2014} and the eventual observation of multifractality~\cite{Rodriguez10}. Last, the scheme could be implemented in a ``reversed way", as proposed in Refs.~\cite{Dao07,pezzeprl2011}, where the ultracold atomic sample under investigation is in the disorder-sensitive state while the resonant transfer is driven to the ``free" state.  This configuration could be used to probe the complex excitation spectra of interacting and disordered quantum gases~\cite{Dao07}, for instance to reveal the predicted gapless excitation spectrum in the Bose glass phase~\cite{Giamarchi1988,Fisher1989}.

\begin{acknowledgments}
This work was supported by ERC (Advanced Grant ``Quantatop''), the Institut Universitaire de France, the Region Ile-de-France in the framework of DIM Nano-K (project QUGASP), the French Ministry of Research and Technology (ANRT) through a CIFRE/DGA grant for V.~D., and the EU-H2020 research and innovation program (Grant No. 641122-QUIC and Marie Sk\l odowska-Curie Grant No. 655933). The authors were granted access to the HPC resources of TGCC under the allocation 2016-057644 made by GENCI (``Grand Equipement National de Calcul Intensif'').
\end{acknowledgments}

V.V. and M.P. contributed equally to this work.

\end{document}


\title{Measurement of spectral functions of ultracold atoms in disordered potentials -- Supplemental Material}

\author{Valentin V. Volchkov}
\affiliation{Laboratoire Charles Fabry, Institut d'Optique, CNRS, 2 Avenue Augustin Fresnel 91127 Palaiseau cedex, France}
\affiliation{Max-Planck-Institute for Intelligent Systems, Spemannstrasse 34, 72076 T\"ubingen, Germany}

\author{Michael Pasek}
\altaffiliation[Present address: ]{Coll\`ege de France, 11 place Marcelin Berthelot, 75005 Paris, France}
\affiliation{Laboratoire Charles Fabry, Institut d'Optique, CNRS, 2 Avenue Augustin Fresnel 91127 Palaiseau cedex, France}
\affiliation{Laboratoire Kastler Brossel, UPMC-Sorbonne Universit\'es, CNRS, ENS-PSL Research University, Coll\`{e}ge de France, 4 Place Jussieu, 75005 Paris, France}

\author{Vincent Denechaud}
\affiliation{Laboratoire Charles Fabry, Institut d'Optique, CNRS, 2 Avenue Augustin Fresnel 91127 Palaiseau cedex, France}
\affiliation{SAFRAN Sensing Solutions, Safran Tech, Rue des Jeunes Bois, Ch\^{a}teaufort CS 80112, 78772 Magny-les-Hameaux, France}

\author{Musawwadah Mukhtar}
\affiliation{Laboratoire Charles Fabry, Institut d'Optique, CNRS, 2 Avenue Augustin Fresnel 91127 Palaiseau cedex, France}

\author{Alain Aspect}
\affiliation{Laboratoire Charles Fabry, Institut d'Optique, CNRS, 2 Avenue Augustin Fresnel 91127 Palaiseau cedex, France}

\author{Dominique Delande}
\affiliation{Laboratoire Kastler Brossel, UPMC-Sorbonne Universit\'es, CNRS, ENS-PSL Research University, Coll\`{e}ge de France, 4 Place Jussieu, 75005 Paris, France}

\author{Vincent Josse}
\email[Corresponding author: ]{vincent.josse@institutoptique.fr}

\affiliation{Laboratoire Charles Fabry, Institut d'Optique, CNRS, 2 Avenue Augustin Fresnel 91127 Palaiseau cedex, France}

\maketitle

\section*{rf-spectroscopy protocol : Theoretical link between the spectral function $\AEk$ and the transfer rate $\Gamma$.}

The aim of this section is to give the explicit derivation of Eq.(1) of the main text. We start with the general expression of the spectral function for non-interacting disordered systems. Second, we derive the expression of transfer rate following the Fermi golden rule and discuss the validity of the approach. Last, we discuss the effect of the inter-atomic interactions in the mean-field regime.

\subsection{Theoretical expression for $\AEk$.}

The spectral function for the energy $\mE_\delta$ can be generally expressed as \cite{bruus}:

\begin{equation}
     \AEkd = - \frac{1}{\pi}\,\mathrm{Im}\, \overline{G}(\mE_\delta,\mathbf{k})=  \overline{\bra{\mathbf{k}}  \delta ( \mE_\delta - H )  \ket{\mathbf{k}}}
 \label{Ak_def_general}
\end{equation}
where $\overline{\dotsb}$ denotes the averaging over disorder realizations, ${G}$ is the retarded Green's function and $H=\bp^2/2m +V(\br)$ is the single particle Hamiltonian of the system. If one introduces the eigenstates $\ket{\psi_\alpha}$ of the Hamiltonian $H$ at energy $\mathcal{E}_\alpha$, Eq.~(\ref{Ak_def_general}) can be rewritten as:  
\begin{equation}
      \AEkd =   \overline{\sum_\alpha |\langle \mathbf{k} | \psi_\alpha \rangle |^2\ \delta(\mE_\delta-\mE_\alpha)} 
 \label{Ak_sum_eigenstates}
\end{equation}
For a single realization of the disorder $V(\br)$, the overlap function $| \langle \mathbf{k} | \psi_\alpha \rangle |^2$ exhibits large fluctuations with the energy  $\mE_\alpha$. However, once disorder averaging is performed, the fluctuations vanish and the overlap function becomes a smooth function. One can then factor the averaged coupling $\overline{ |\langle \mathbf{k} | \psi_\delta \rangle |^2}$ at the energy $\mE_\delta$ from the sum, yielding:
\begin{eqnarray}
 \AEkd & \sim & \overline{ |\langle \mathbf{k} | \psi_\delta  \rangle|^2} \cdot \overline{\sum_\alpha \delta(\mE_\delta-\mE_\alpha)} \nonumber \\
& \sim &  \overline{ |\langle \mathbf{k} | \psi_\delta  \rangle |^2}  \cdot \rho (\mathcal{E}_\delta)\; , \label{Eq_Ak_rho}
\end{eqnarray}
where $  \rho(\mE)$ is the disorder averaged density of states.

\begin{figure}
\includegraphics[width=0.9\columnwidth]{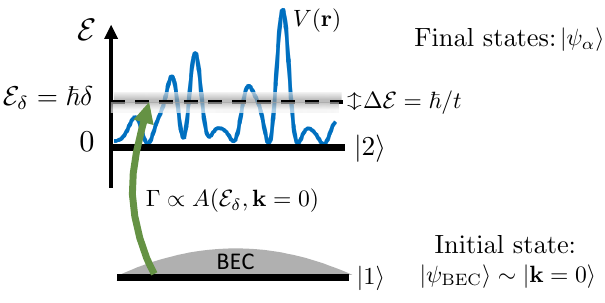}
\caption{\textbf{Measurement scheme of the spectral function}. The initial state is a BEC created in the disorder insensitive internal state $\ket{1}$. Its spatial wavefunction is approximately $\ket{\bk=0}$. A weak rf field drives the transition from the initial state to the internal final state $\ket{2}$ that is sensitive to the disordered potential $V(\br)$, with a transfer rate $\Gamma$. The spatial wavefunctions of the final states are denoted $\psi_\alpha$. These targeted states have an energy centered around the resonant condition $\mathcal{E}_\delta=\hbar \delta$, where $\delta$ is the rf-detuning with respect to the bare transition frequency between states $\ket{1}$ and $\ket{2}$. The energy spread around $\mathcal{E}_\delta$ is Fourier limited by the duration $t$ of the coupling: $\Delta \mathcal{E}=\hbar/t$.   }
\label{figSupp_scheme}
\end{figure}

\subsection{Expression of the transfer rate $\Gamma$}

For clarity, the rf-transfer scheme presented in the letter is reproduced here in figure~\ref{figSupp_scheme}, with some additional details. We consider the simple case where the two internal states $\ket{1}$ and $\ket{2}$ are directly coupled by an oscillating rf magnetic field at frequency $\omega_\mathrm{rf} $. The two-photon rf transition used in the experiment (see Fig.~\ref{fig:freeRes}) is described in the specific section ``Rf-spectroscopy protocol: Experimental details" below. The magnetic dipolar interaction is $\hat{W}= \hat{W}_0 \,e^{-i\omega_\mathrm{rf} t}\, +\, \mathrm{c.c}$, where $\hat{W}_0= \hbar \Omega \, |2\rangle\langle1| $ is the static coupling and $\Omega $ the Rabi frequency.

A key feature is that  the energy spectrum of the targeted states in the disordered potential can be considered as a continuum, due to the finite energy resolution $\Delta \mE=\hbar/t$, where $t$ is the duration of the rf pulse (see Fig.~\ref{figSupp_scheme}). If the Rabi frequency $\Omega$ is weak enough, the transfer rate can be expressed by the Fermi golden rule (see e.g.~Ref.~\cite{grynberg2010introduction}): 
\begin{equation}
\Gamma=\frac{2\pi}{\hbar} \sum_{f}|\braket{f |  \hat{W}_0 | i  }|^2\;\delta_\mathrm{t}(E_f- E_i -\hbar \omega_\mathrm{rf}).\label{Eq_rof}
\end{equation}
Here $\ket{i}=\ket{1}\ket{\psi_\mathrm{BEC}}$ is the initial state of \emph{ total} energy $E_i$ and $\ket{f}=\ket{2}\ket{\psi_\alpha}$ refers to the final states of \emph{ total }energy $E_f$. 
The function $\delta_t (E)=2\hbar\sin^2(Et/2\hbar)/\pi t E^2$ is a narrow function of width $\Delta \mE=\hbar/t$ [see Fig.~(\ref{figSupp_scheme})]. 
For long durations, it behaves essentially as a Dirac function, with $\lim_{t\rightarrow+\infty} \delta_t(E) = \delta(E)$. Since the coupling between the initial and final states writes $\braket{f |  \hat{W}_0 | i  }=\hbar \Omega \cdot \braket{\psi_\alpha | \psiBEC}$, Eq.~(\ref{Eq_rof}) can be directly simplified as:
\begin{equation}
\Gamma(\delta)=2\pi \hbar \Omega^2 \sum_{\alpha}|\braket{\psi_\alpha |\psiBEC  }|^2\;\delta_\mathrm{t}(\mE_\alpha-\mE_\delta),\label{Eq_rof_overlapp}
\end{equation}
where we used the relation $E_f- E_i -\hbar \omega_\mathrm{rf}=\mE_\alpha-\mE_\delta$, with $\mE_\delta=\hbar \delta$ is the final \emph{external} energy targeted by the rf frequency (see fig.~\ref{figSupp_scheme}).

In order to make the link with the expression~(\ref{Eq_Ak_rho}) of the spectral functions, we use two assumptions. First, we identify the BEC's wavefunction with a purely uniform one, i.e. $\ket{\psiBEC}\sim\ket{\bk=0}$. This assumption is meaningful since the BEC coherence length is much longer than the correlation length $\sigma$ of the disordered potential. 
Second, we use an ``ergodic'' hypothesis, i.e. replace the averaging over states within the band of width $\Delta \mE$ by an averaging over different realizations of the disordered potential.  
As for the ensemble averaging, the energy averaging smoothes the overlap function $|\braket{\psi_\alpha | \bk=0 }|^2$, which becomes a slowly varying function at the energy scale given by the resolution. One can thus replace the discrete sum by an integral involving the averaged density of states $\rho(\mE)$~\footnote{Note that additional disorder averaging could be performed in the experiment. However, the energy averaging for a single disorder realization already reduces fluctuations considerably, and no residual fluctuations were observed in the experiment.}. Altogether we obtain:
 \begin{equation}
\Gamma(\delta)=2\pi \hbar \Omega^2   \; \overline{|\braket{\psi_\delta | \bk=0  }|^2}\; \rho(\mE_\delta) \propto \AEkzd \; .\label{Eq_Gamma_Aek}
\end{equation}

The derivation of Eq.~(\ref{Eq_Gamma_Aek}) relies on the assumption that the overlap function $\overline{|\braket{\psi_\delta |\bk=0 }|^2}$ is a slowly varying function of the energy. The characteristic energy scale  $\hbar \Delta_\mathrm{w}$ of variation of this overlap function is the energy width of the continuum that is coupled to the initial state. In our specific case, it is nothing but the width of the spectral function $\AEkz.$
The generic case of an initial state coupled to a continuum of finite width has been studied in great details in the literature, see e.g.  Ref.~\cite{grynberg2010introduction}. 
The Fermi golden rule is valid provided that the Rabi frequency $ \Omega $ remains much smaller than the continuum width, i.e.:
\begin{equation} 
\Omega \ll \Delta_\mathrm{w}\; .
\label{Eq_crit_Rabi}
\end{equation} 
Otherwise the system cannot differentiate the continuum from a genuine discrete two level system, and Rabi oscillations take place. In all cases, the condition $\Omega\ll \Delta_\mathrm{w}$ has been carefully checked in the experiment. In particular we verified that the shape of the measured spectral function is not modified when $\Omega$ is varied.
 
 Another parameter is the duration $t$ of the rf pulse. Here we choose $t$ long enough for the energy resolution $\Delta \mE$ to be much smaller that the width of the spectral function, and short enough to deplete only a very small fraction of the initial state (typically a few percent). Altogether, we operate in the regime:
 \begin{equation} 
 \hbar/\Delta_\mathrm{w} \ll t \ll 1/\Gamma
\label{Eq_crit_duration}
\end{equation} 

\subsection*{Raman \emph{v}s rf coupling scheme}

In the protocol described in figure~\ref{figSupp_scheme}, we measure the spectral functions at $\bk=0$ because the rf field carries a negligible momentum. The scheme can be adapted to measure the spectral function with a finite momentum $\bk$ by using a two-photon Raman transition instead (see e.g.~\cite{Dao07,Dao09}). In that case, a net momentum of $\mathbf{\Delta k}=\mathbf{k}_2-\mathbf{k}_1$, where $\mathbf{k}_{1,2}$ are the respective wave numbers of the Raman beams, is transferred to the atoms. By varying the angle between the two lasers, the momentum transfer can be tuned between $0$ and $2 k_\mathrm{L}$, where $k_1\sim k_2\equiv k_\mathrm{L}$.
Formally, the coupling matrix element writes now as $\braket{f | \hat{W}_0 | i  }=\hbar \Omega_\mathrm{eff} \cdot \braket{\psi_\alpha| e^{i \mathbf{\Delta k}\cdot \br} | \psiBEC}$, where $\Omega_\mathrm{eff}$ is the effective Rabi coupling. Taking once again $\ket{\psiBEC}\sim \ket{\bk=0}$, one has $\braket{f |  \hat{W}_0 | i  }=\hbar \Omega_\mathrm{eff} \cdot \braket{\psi_\alpha |\mathbf{\Delta k}}$ and $\Gamma \propto A(\mathbf{\Delta k},\mE)$.

\subsection*{Effect of mean-field interactions and the harmonic trap}

The above derivation of Eq.~(\ref{Eq_Gamma_Aek}) can be extended to interacting particles, provided that the interactions can be treated at the mean-field level (i.e. for dilute atomic samples). In that case, the interactions play the role of an effective potential that should be taken into account in the energy resonance condition $E_f=E_i+\hbar \omega_\mathrm{rf}$.

The short range interactions depend on the atomic internal states $(i,j)$. They are fully characterized by the scattering lengths $a_{ij}$, or equivalently by the interaction parameters $g_{ij}=4\pi\hbar^2 a_{ij}/m$. More precisely, the mean-field interaction energy for one particle immersed in the BEC writes $\mE_{i,\mathrm{mf}}=g_{11} \, n_\mathrm{BEC}(\br)$ where $n_\mathrm{BEC}(\br)$ is the BEC density. Once the particle has been transferred in state $\ket{2}$, it experiences the mean-field $1-2$ interaction with the remaining atoms in the BEC, together with the $2-2$  with the atoms already transferred. For weak rf coupling, the atomic density in state $\ket{2}$ is very low and the latter interaction can be neglected. Since $a_{11}\simeq a_{12}$ for $^{87}$Rb atoms~\cite{Egorov13}, the mean-field interaction energy is the same in states  $\ket{1}$ and  $\ket{2}$ and cancels out in the energy resonance condition. Similarly, the shallow harmonic trapping potential is equal in both states and also cancels out.

\section*{Rf-spectroscopy protocol: Experimental details}

The following subsections give experimental details on the BEC parameters, the implementation of the two-photon rf transition used in the experiments and the transfer energy resolution. 

\subsection*{Preparation of the BEC}

The BEC is created starting from $^{87}$Rb atoms in the hyperfine state $\ket{1}\equiv\ket{F=1,m_F=-1}$ that are confined in a crossed optical dipole trap formed by two orthogonal laser beams at a wavelength of $1064\,\mathrm{nm}$ (as described in Ref.~\onlinecite{jendrzejewski2012b}). In addition, a magnetic field gradient $B'=30.46\,\mathrm{G}.\mathrm{cm}^{-1}$ is applied in the vertical direction in order to counteract gravity. We use forced evaporation to obtain an almost pure and dilute BEC with around $2\times 10^5$ atoms, the trapping frequencies being around $\omega_\mathrm{T}/2\pi\sim10\,\textrm{Hz}$ in all directions. The BEC's chemical potential is $\mu_\mathrm{BEC}=g_{11}\, n_\mathrm{BEC}(\br=0)\sim h \times 140$~Hz (see section ``Effect of mean-field atomic interaction" above), and the Thomas-Fermi (TF) radius is $R_\mathrm{TF}=\sqrt{2\mu_\mathrm{BEC}/ m \omega_\mathrm{T}^2}\;\sim18\,\mu\textrm{m}$. 

The BEC's size $R_\mathrm{TF}$ is much larger than the typical correlation length $\sigma$ of the laser speckle potential (see section `` Statistical properties : measurement of the spatial auto-correlation function" below).

\begin{figure}
\includegraphics[width=0.99\columnwidth]{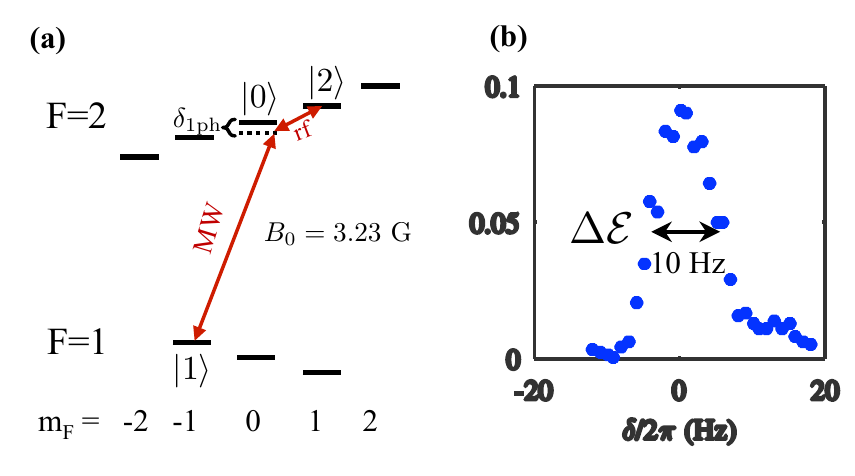}
\caption{\textbf{Two-photon transition scheme.} (a) Two-photon transition between the ``clock states'' $\ket{1}$ and $\ket{2}$ (via the intermediate state $\ket{0}$) of the ground state hyperfine structure of $^{87}$Rb. The two-photon transition uses a microwave magnetic field at frequency $\omega_\mathrm{MW}$ and a rf magnetic field at frequency $\omega_\mathrm{rf}$.  (b) Two-photon resonance in the absence of disordered potential for a coupling duration $t=100$~ms. The resonance curve has a Fourier limited width $\Delta \mE/h\sim 10$~Hz. }
\label{fig:freeRes}
\end{figure}

\subsection*{Two-photon rf transition}

The hyperfine ``clock states'' $\ket{1}\equiv\ket{F=1,m_F=-1}$ and $\ket{2}\equiv\ket{F=2,m_F=1}$ used in the experiment are shown in Fig.~\ref{fig:freeRes}(a). We suppressed the broadening effect of magnetic field fluctuations by working at a bias field $B=3.23\,\textrm{G}$, for which the magnetic moments of $\ket{1}$ and $\ket{2}$ are equal \cite{Harber02}.

Due to selection rules (the angular momentum difference is $\Delta m_F=2$), these two states can only be coupled using a two-photon rf transition via the intermediate state $\ket{0}\equiv\ket{F=2,m_F=0}$. This ``two-photon'' transition is realized with (i) a microwave magnetic field at frequency $\omega_{\textrm{MW}}/2\pi \sim 6831.9\,\textrm{MHz}$ that drives the $\ket{1}\rightarrow\ket{0}$ transition and (ii)  a rf field at frequency $\omega_{\textrm{rf}}/2\pi\sim 2.8\,\textrm{MHz}$ that drives the $\ket{0}\rightarrow \ket{2} $ transition.

The single photon transition to the intermediate state $\ket{0}$ was detuned by $\delta_{\mathrm{1ph}} / 2\pi\sim 0.5$~MHz. In that case, the entire system can be seen as an effective 2-level system, with an effective Rabi frequency $\Omega= \Omega_{\mathrm{MW}}\Omega_{\mathrm{rf}}/\delta_{\mathrm{1ph}}$, where $\Omega_{\mathrm{MW}}$ and $\Omega_{\mathrm{rf}}$ are the Rabi frequencies associated to each transitions.  $\Omega$ was measured by observing two-photon Rabi-oscillations from $\ket{1}$ to $\ket{2}$ in the absence of disordered potential. In practice, we adjusted the coupling $\Omega$ and the duration $t$ of the transfer for each disorder amplitude $V_0$ in order to fulfill the criteria~(\ref{Eq_crit_Rabi}) and~(\ref{Eq_crit_duration}). Consistently, the maximum number of transferred atoms stayed fairly small in these conditions (on the order of $5\times10^3$ atoms, i.e.~a few percent of the atoms in the initial BEC). Note that the effective Rabi frequency ranged up at maximum to $\Omega/ 2\pi\sim 80$~Hz, i.e.~much lower than the single photon detuning $\delta_{\mathrm{1ph}}$.

\subsection*{Transfer resolution}

The transfer resolution of the two-photon rf transition results from the convolution of the resolution associated with the transfer time, i.e.  $  \Delta \mathcal{E}=\hbar / t$. 
Additional sources of broadening are estimated to be negligible:
(i) The broadening due to the magnetic field gradient across the size of the condensate is estimated of the order of $1\,\textrm{Hz}$ (ii) The residual broadening due to difference of mean-field interaction between the excited $\ket{2}$ and initial $\ket{1}$ state is 
of the order of $(g_{12}-g_{11})n_\mathrm{BEC}(\br)\sim (1-a_{12}/a_{11})\;\mu_\mathrm{BEC}/h$ (see section ``Effect of mean-field atomic interaction" above). Since $a_{12}/a_{11}\approx0.976$ \cite{Egorov13}, an upper bound is $\sim 3$~Hz.

 Altogether, the transfer resolution is then Fourier limited for our experimental conditions. In practice, we adapted the resolution to the typical energy span of the spectral function for each disorder amplitude $V_0$, so that it does not affect the observed profile. More precisely, the duration was varied from $t=5$~ms ($\Delta \mE/h =200$~Hz for the strongest disorder amplitude $V_0/h\sim4$~kHz) to $t=100$~ms at maximum ($\Delta \mE/h =10$~Hz for the lowest disorder $V_0/h\sim60$~Hz). As an example, Fig.~\ref{fig:freeRes} shows the two-photon resonance in the absence of disordered potential for the largest duration time, i.e. $t=100$~ms.

\vspace{3ex}
\section*{Laser speckle disordered potential: experimental characterization and numerical simulation}

\subsection*{State dependent disordered potential}

In order to obtain a state dependent disordered potential we used a laser speckle tuned close to the optical transition   $F=2 \rightarrow F'=3$ (see Figs.~2 and 1(a) in the main article), with a detuning $\Delta_\mathrm{L}$ much smaller than the hyperfine splitting of the ground state separating the ``clock states'' $|\Delta|\ll\Delta_{\textrm{HFS}}$ (similar to the ``tune-in'' scheme as described by \cite{leblanc07}).  
With this configuration, the selectiveness $V_{\ket{2}}/V_{\ket{1}}$  of the disorder potential for $\ket{2}$ is roughly given by $\Delta_{\textrm{HFS}}/|\Delta|\approx 100$.

\begin{figure}
\includegraphics[width=0.99\columnwidth]{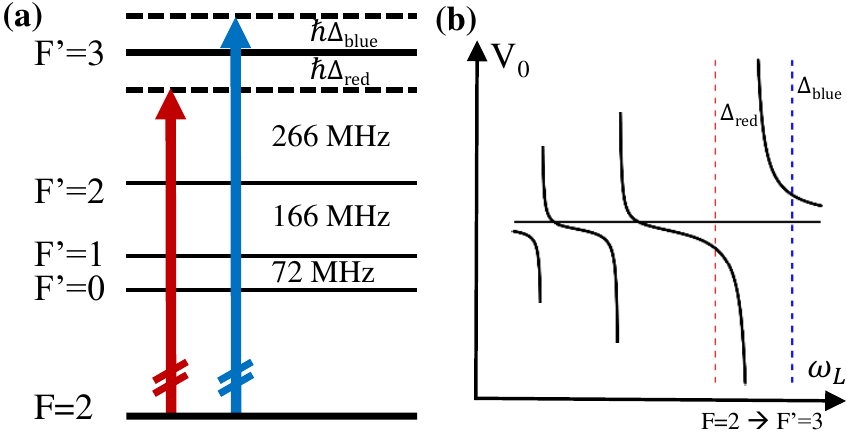}
\caption{\textbf{Amplitude of the disordered potential.} (a) Hyperfine structure of the excited state and detunings of the laser speckle ($\Delta_\mathrm{red}$ and $\Delta_\mathrm{blue}$ for the red-detuned and blue-detuned case respectively). (b) Disorder amplitude $V_0$ in state $\ket{2}$ as a function of the speckle laser frequency around the $F=2 \rightarrow F'=3$ resonance.}
\label{fig:excState}
\end{figure}

Being close to the resonance, the precise hyperfine structure of the excited state has to be taken into account, as shown in panel (a) of Fig.~\ref{fig:excState} of the present supplemental material. Furthermore, the speckle laser was frequency offset-locked to a saturation spectroscopy stabilized laser, allowing us to control its detuning with an accuracy of $1\,\mathrm{MHz}$. The exact detuning for the attractive and repulsive disordered potentials was calibrated by performing two-photon rf-spectroscopy of atoms in homogeneous light field (instead of a speckle field) produced by the same laser. The absolute light shifts created by the red and blue detuned light were adjusted to be equal.
The experimentally determined values for $\Delta_{\textrm{blue}}=2\pi\times81\,\textrm{MHz}$ and $\Delta_{\textrm{red}}=-2\pi\times 73\,\textrm{MHz}$ are in agreement with an optical dipole potential model taking all excited hyperfine states into account [see Fig.~ \ref{fig:excState}(b)].

In order to calibrate precisely the disorder amplitude $V_0$, we took benefit of the excellent agreement between experiments and numerics. The calibration was done by adjusting our measured spectral function with the numerically-computed spectral function at $V_0/h=-121\,\textrm{Hz}$ (attractive laser speckle disorder). It leads to a calibration of the disorder amplitude with a 4$\%$ precision. Note that this calibration corresponds to a global correction of about 14$\%$ compared to an independent calibration based on photometric measurements. Such deviation is not surprising (it is comparable to the one found in~\cite{semeghini2014}) since the light intensity field cannot be measured at the location of the atoms (experiments take place inside a vacuum chamber). Here we estimate that the uncertainty mainly originates from a slight mis-estimation of the spatial extension of the laser speckle field.

In practice, the laser speckle power was varied between $0.1$ and $10\,\mu\textrm{W}$ to vary the disorder amplitude from $V_0/h\sim60$~Hz to $V_0/h\sim4$~kHz (see Fig.~2 of the main text). Despite the proximity to the resonance, no heating or losses due to inelastic scattering were observed at the short time scales considered in the experiments.

\subsection*{Statistical properties of the disordered potential: measurement of the spatial auto-correlation function}

The laser speckle is created by passing a laser beam through a diffusive plate.  As illustrated in Fig.~\ref{fig:corr}(a), the incoming wave that illuminates the diffusive plate is converging around the position $d$ of the atoms, such that they experience the Fraunhofer's diffraction pattern of ground plate (i.e.~the so-called Fourier speckle configuration~\cite{goodman2007}). The intensity profile of the illumination on the diffusive plate is a Gaussian shape, of waist $w$  (radius at $1/e^2$), truncated by a circular diaphragm of diameter $D$. The estimated geometrical parameters are: $D=20.3(1)$~mm, $w=9(1)$~mm and $d=15.2(5) $~mm. The diaphragm sets the maximal numerical aperture to $\mathrm{NA}=\sin(\theta_\mathrm{max})=0.55(2)$.

 In order to characterize the laser speckle field, the random intensity pattern was recorded at the position of the atoms with an high-resolution optical microscope, see Fig.~\ref{fig:corr}(b). The measurement was done \emph{ex-situ}, i.e. outside the vacuum chamber, but with an exact replica of the geometrical configuration. The extracted two-point correlation functions in the transverse ($z$ direction, same along $y$) and longitudinal directions ($x$ direction) are shown as blue squares in Fig.~\ref{fig:corr}(c), together with the half-width-at-half-maximum (HWHM) lengths. They correspond to $\mathrm{HWHM}_\perp \approx 0.42(1)\, \mu\mathrm{m}$ and $\mathrm{HWHM}_\parallel \approx 2.05(5)\,\mu\mathrm{m}$. More details about the experimental configuration and the laser speckle characterization can be found in Ref.~\onlinecite{richard15}.

For consistency with previous theoretical work~\cite{pasek2016anderson}, we define the correlation lengths of the speckle potential as $\sigma_{\parallel,\perp} = \mathrm{HWHM}_{\parallel,\perp} / 1.39156$,
which yields $\sigma_\perp \approx 0.306\, \mu\mathrm{m}$ and $\sigma_\parallel \approx 1.45\, \mu\mathrm{m}$. These values are used for the calculation of the correlation energy $\mathcal{E}_\sigma = \hbar^2/m(\sigma_\perp^2 \sigma_\parallel)^{2/3}\sim441~\mathrm{Hz} $ (see main text).

\bigskip

\subsection*{Numerical simulation of the laser speckle}

\begin{figure}[b!]
\includegraphics[width=1\columnwidth]{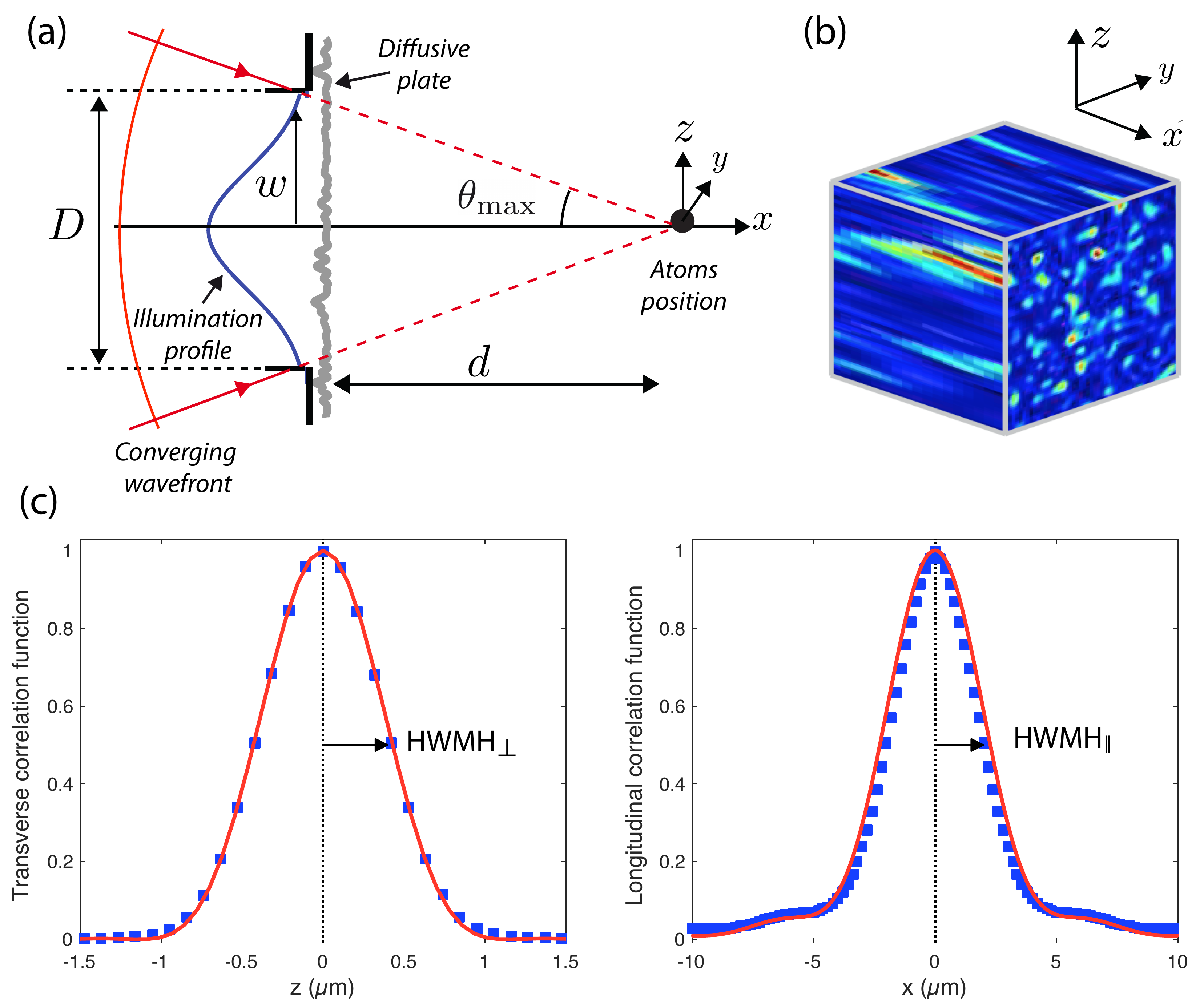}
\caption{\textbf{Laser speckle characterization.} (a) Schematic representation of the experimental configuration used for the speckle generation, with $\mathrm{NA}=\sin(\theta_\mathrm{max})=0.55(2)$ (see text). (b) 3D representation of the experimental laser speckle field recorded with an high aperture microscope at the position of the atoms.  (c) Transverse  and longitudinal correlation functions of the laser speckle, respectively along the $z$ (same along $y$) and $x$ directions. Blue squares: experimental autocorrelation functions extracted from the measurement shown in (b). Red solid lines: autocorrelation function for numerically simulated laser speckle [following Eqs.~(\ref{supsingle}) and~(\ref{sup2})]. Horizontal black arrows indicates the half-maxima $\mathrm{HWHM}_\perp=0.42(1)$~$\mu$m and $\mathrm{HWHM}_\parallel=2.05(5)$~$\mu$m of the correlation function.}
\label{fig:corr}
\end{figure}

We designed the numerical disorder to reproduce the experimental geometry represented in Fig.~\ref{fig:corr}(a). In particular, the numerical speckle generation takes into account the presence of a diaphragm, and hence deviates from the usual pure Gaussian and Lorentzian character (see e.g.~\cite{goodman2007}) for the two-point correlation function of the intensity in the transverse and longitudinal directions [see Fig.~\ref{fig:corr}(c)]. 
\begin{widetext}
To do so, the generation of the numerical disorder was done through the use of a phase mask:
\begin{equation}\label{supsingle}
\mathcal{P} (\mathbf k) =\delta(|\mathbf{k}|-k_L) \exp\left[-\frac{(\tan\theta)^2}{(w/d)^2}\right]  \Theta \left[  \frac{D}{2 d} - \tan|\theta|   \right],
\end{equation}
where $k_\mathrm{L}$ is the wavenumber of the monochromatic laser beam, $\theta \in [-\pi/2,\pi/2]$ the angle between the wavevector $\mathbf{k}$ and the beam axis, $w$ the beam waist, $D$ the diaphragm diameter and $d$ the position where the illumination field converges.
\end{widetext}
The numerical potential was then generated starting from a local random complex field $E_{u}(\mathbf r)$, whose real and imaginary parts are uncorrelated Gaussian random variables.
The convolution of this random field with the phase mask in Eq.~(\ref{supsingle}) yields the proper statistical distribution of the amplitude of the electric field~\cite{goodman2007}
\begin{equation}\label{sup2}
E(\mathbf r)= \sum_{\mathbf k}E_u(\mathbf k) \mathcal{P} (\mathbf k) e^{i \mathbf k \cdot \mathbf r},
\end{equation}
which reproduces the coherent superposition of the various plane waves scattered by the diffusive plate. 
Parameters of the numerical phase mask, Eq.~(\ref{supsingle}), that yield the best agreement with measured two-point correlation functions of the speckle intensity, see Fig.~\ref{fig:corr}(b), are $D = 20.4\,\mathrm{mm}$, $d = 15.2\,\mathrm{mm}$ and $w = 9.9\,\mathrm{mm}$, in excellent agreement with the experimentally-inferred values. 

\section*{Numerical computation of the spectral function}

\subsection*{Method}

We start with the definition, Eq.~(\ref{Ak_def_general}), of the spectral function $A(\mathcal{E},\mathbf{k})  =  \overline{\bra{\mathbf{k}}  \delta ( \mathcal{E} - H ) |\mathbf k \rangle} =  - \frac{1}{\pi}\,\mathrm{Im}\, \overline{G}(\mE_\delta,\mathbf{k})$. The spectral function can be evaluated numerically in many ways. The most common in the literature starts from the spectral, so-called Lehmann, representation of the Green's function (as used in, e.g., Ref.~\onlinecite{semeghini2014}). We choose to use instead the temporal representation of the Green's function:
\begin{equation}
\label{Ak_temp}
A(\mathcal{E},\mathbf{k}) = \frac{1}{\pi\hbar}\,\mathrm{Re}\,\int_0^{\infty} \overline{\langle \mathbf{k}|\rme^{-\rmi H t/\hbar}|\mathbf k \rangle}  \rme^{\rmi \mathcal{E}t/\hbar}\ \rmd t.
\end{equation}
The numerical computation of the spectral function $A(\mathcal{E},\mathbf{k})$ thus amounts to: (1) Computing the overlap between an initial plane-wave excitation $\ket{\mathbf{k}}$ 
and its time-evolution under the (disordered) Hamiltonian $H$; (2) Averaging this overlap over multiple disorder configurations (typically between 2000 and 4000 realizations depending on the disorder amplitude); (3) Using the Fourier transform to go from time- to energy-representation.

For the time-propagation algorithm, we use an iterative scheme based on the expansion of the evolution operator $\exp (-i H\Delta t /\hbar)$ over a time-step $\Delta t$ in a series of Chebyshev polynomials of the Hamiltonian \cite{roche1997,fehske2009,trappe2015}.
In order to check the numerical accuracy of our results, we verify that they satisfy the first three sum rules of the spectral function, namely \cite{trappe2015}:
\begin{align}
\int \rmd \mathcal{E}\, A_\mathbf{k} (\mathcal{E}) &= 1,\\
\int \rmd \mathcal{E}\, \mathcal{E} A_\mathbf{k} (\mathcal{E}) &= \frac{\hbar^2\mathbf{k}^2}{2m} + \overline{V},\\
\int \rmd \mathcal{E}\, \mathcal{E}^2 A_\mathbf{k} (\mathcal{E}) &= \left( \frac{\hbar^2\mathbf{k}^2}{2m} + \overline{V} \right)^2 + \overline{\delta V^2},
\end{align}
where $\overline{\delta V^2}=V_0^2$ is the variance of the speckle potential distribution.

\subsection*{Effect of the residual disorder in state $\ket{1}$}

In the experiment, the residual disorder on state $\ket{1}$ is of the order of $0.01|V_0|$ (see section ``State dependent disordered potential'' above). 
For the largest value of $|V_0|$ ($\approx h\times 4\,\mathrm{kHz}$), it results in a potential around 40 Hz on the state $\ket{1}$ that perturbs the initial BEC (of chemical potential $\mu_\mathrm{BEC}/h\sim 140 $~Hz), and consequently the transfer $|1\rangle \to |2\rangle$. This effect can be approximately taken into account in the numerical calculation of the spectral function.
To do so, we performed a preliminary propagation of the initial $\ket{\mathbf{k}=0}$ state in the presence of the same disorder realization but with residual amplitude $V_{\ket{1}}=-0.01|V_0|$ (which is always blue-detuned since $\Delta_\mathrm{HFS}\gg\Delta_\mathrm{L}$, see Fig.~1 of the main text), before performing the time-evolution under the disordered Hamiltonian for atoms in state $\ket{2}$ [i.e. step (1) in the above-described numerical procedure]. 
The duration of this preliminary propagation was taken to half the experimental transfer time, though different durations had little influence on the resulting spectral function. This is because the product $|V_{\ket{1}}|t/h$ is at maximum of the order of $0.2,$ so that the state $\ket{1}$ is only slightly distorted by the residual disorder over the duration of the experiment.  The resulting spectral functions are displayed as solid brown lines in Panels (I.f) and (II.f) of Fig.~2 of the main text.